\documentclass[10pt,final, twocolumn]{IEEEtran}

\usepackage{times}
\usepackage{amsmath,dsfont}
\usepackage{amssymb,amsthm}
\usepackage{epsfig,verbatim}
\usepackage{subfigure}
\usepackage{setspace}
\usepackage{color}
\usepackage{cite}
\usepackage{epstopdf}
\usepackage{graphics}
\usepackage{accents}
\usepackage{acronym}

\usepackage{hyperref}
\hypersetup{colorlinks}
\usepackage{booktabs}
\usepackage{mathtools}
\usepackage{algorithm}
\usepackage{algorithmicx}
\usepackage{algpseudocode}
\usepackage{amsmath}
\usepackage{enumitem}

\newtheorem{theorem}{Theorem}

\acrodef{adc}[ADC]{analog-to-digital convertor}
\acrodef{cs}[CS]{compressed sensing}
\acrodef{dtft}[DTFT]{discrete-time Fourier transform}
\acrodef{dnn}[DNN]{deep neural network} 
\acrodef{csi}[CSI]{channel state information}
\acrodef{map}[MAP]{maximum a-posteriori probability}
\acrodef{snr}[SNR]{signal-to-noise ratio}
\acrodef{bs}[BS]{base station} 
\acrodef{iot}[IOT]{Interent of Things}
\acrodef{mimo}[MIMO]{multiple-input multiple-output}
\acrodef{mse}[MSE]{mean-squared error}
\acrodef{mmse}[MMSE]{minimum mean-squared error}
\acrodef{pdf}[PDF]{probability density function}
\acrodef{rv}[RV]{random variable}
\acrodef{fec}[FEC]{forward error correction}
\acrodef{dma}[DMA]{dynamic metasurface antenna}
\acrodef{lti}[LTI]{linear time-invariant}
\acrodef{wss}[WSS]{wide-sense stationary}
\acrodef{psd}[PSD]{power spectral density}
\acrodef{ser}[SER]{symbol error rate} 
\acrodef{ber}[BER]{bit error rate} 
\acrodef{sgd}[SGD]{stochastic gradient descent} 
\acrodef{isi}[ISI]{intersymbol interference}  
\acrodef{awgn}[AWGN]{additive white Gaussian noise} 
\acrodef{ut}[UT]{user terminal} 
\acrodef{mmw}[mmWave]{millimeter wave}

\IEEEoverridecommandlockouts

\title{ 

Illumination Design for Joint Imaging and Wireless Power Transfer Systems}
\author{\IEEEauthorblockN{Qianyu Yang,~\IEEEmembership{Student Member,~IEEE},  Haiyang Zhang,~\IEEEmembership{Member,~IEEE}, Chunguo Li,~\IEEEmembership{Senior Member,~IEEE},\\  Ruiqi Liu,~\IEEEmembership{Member,~IEEE}, Baoyun Wang,~\IEEEmembership{Senior Member,~IEEE}  \\
	}   
	
	\thanks{ 
	
		}

}

\begin{document}
	
	\maketitle
	\pagestyle{empty}
	\thispagestyle{empty}

\begin{abstract}

This paper presents a novel concept termed Integrated Imaging and Wireless Power Transfer (IWPT), wherein the integration of imaging and wireless power transfer functionalities is achieved on a unified hardware platform. IWPT leverages a transmitting array to efficiently illuminate a specific Region of Interest (ROI), enabling the extraction of ROI's scattering coefficients while concurrently providing wireless power to nearby users. The integration of IWPT offers compelling advantages, including notable reductions in power consumption and spectrum utilization, pivotal for the optimization of future 6G wireless networks. 
As an initial investigation, we explore two antenna architectures: a fully digital array and a digital/analog hybrid array. Our goal is to characterize the fundamental trade-off between imaging and wireless power transfer by optimizing the illumination signal. With imaging operating in the near-field, we formulate the illumination signal design as an optimization problem that minimizes the condition number of the equivalent channel. To address this optimization problem, we propose an semi-definite relaxation-based approach for the fully digital array and an alternating optimization algorithm for the hybrid array. 
Finally, numerical results verify the effectiveness of our proposed solutions and demonstrate the trade-off between imaging and wireless power transfer.


	\end{abstract}

	\section{Introduction}
Leveraging high-frequency bands and extensive antenna arrays is anticipated to empower 6G systems to predominantly function within the radiation near-field region, offering unparalleled communication and sensing capabilities, flexibility, and resolution \cite{6Gcontext,ITU6G}. Within this framework, notable research contributions have underscored the potential of holographic imaging \cite{nearimaging}, a technique widely utilized in various environmental sensing applications such as security inspections and medical diagnoses. Typically integrated into internet of things (IoE) devices for decision-making processes, these applications face limitations imposed by device size and hardware costs, often constrained by battery capacities \cite{energy-save, battery}. Radio frequency-based wireless power transfer (WPT) emerges as a promising technology for wirelessly charging such devices, presenting a viable solution to address energy scarcity concerns \cite{WPT, WPT2}.

Traditionally, imaging and WPT have been explored as separate domains. Imaging tasks have conventionally relied on technologies like microwaves or visible light, including lidars \cite{lidar}, with limited exploration on repurposing wireless communication systems for imaging unfamiliar objects within a designated Region of Interest (ROI) \cite{waveimage,3Dimage}. However, in the context of forthcoming wireless systems, the allure of near-field imaging, particularly through the utilization of extremely large Massive MIMO (XL-MIMO) and high-frequency bands, has garnered attention. Initial investigations have delved into the beamforming challenges intrinsic to this approach \cite{holographic}, hinting at a future where imaging could seamlessly integrate into the IoE sensor network without necessitating a dedicated imaging infrastructure.
On the other hand, the concept of radio wave-powered devices has been a focal point in pioneering discussions \cite{WPT3,SWIPT}. Notably, within the realm of wireless communication, there is a growing interest in amalgamating WPT with existing systems, such as simultaneous wireless information and power transfer (SWIPT) \cite{SWIPT2} and integrated sensing and wireless power transfer (ISWPT) \cite{ISWPT}. These technologies aim to streamline communication and power delivery at the transmitter's end to curtail device expenses. Despite these advancements, the integration of imaging and WPT remains unexplored due to the traditionally distinct nature of these fields and their operational paradigms.

To exploit higher array gain in such scenarios, precoding and combining techniques play a crucial role. The most adaptable solution for an array of radiating elements is the fully digital architecture, where each antenna element connects to a dedicated Radio Frequency (RF) chain, enabling precise beam control. However, the scalability of this approach poses challenges due to the substantial costs associated with expensive RF components and high power consumption \cite{zirtiloglu2022power}. To address the complexity of fully digital architectures, hybrid solutions have emerged, blending traditional antenna technologies with the integration of multiple antennas or panels to an RF chain through dedicated analog circuitry \cite{levy2024rapid}.
These hybrid architectures optimize resources by employing fewer RF chains relative to antenna elements. They achieve this by combining low-dimensional digital processing with high-dimensional analog precoding techniques, such as phase-shifter networks \cite{hybridarray}. Such methodologies find applications in various large-scale antenna technologies, including near-field communications \cite{nearcom} and wideband communications \cite{widecom}.

In this paper, we propose a novel concept that merges imaging and WPT into a unified hardware platform, terming the envisioned scenario as integrated imaging and wireless power transfer (IWPT). The proposed IWPT system is envisioned for deployment in common IoE settings that necessitate imaging functionalities, like security inspections and nondestructive testing, enabling the simultaneous operation of imaging and charging sensors. The integration of imaging and WPT within a single system presents several advantages. Firstly, it streamlines device costs and optimizes spectrum resources compared to conventional separate designs, as WPT can be achieved by utilizing the imaging beams, obviating the necessity for additional power beam configurations. Moreover, the continuous coverage provided by the illuminating beam over the ROI ensures a stable energy supply to devices within this area. Additionally, harvesting energy from the illumination signals enhances the overall energy efficiency of the platform. Furthermore, the proximity of charge sensors to the ROI typically situates them within the radiation near-field range of the transmitter, which amplifies the efficiency of WPT due to the near-field propagation mechanism, thereby offering additional benefits to the system's energy transfer capabilities.

In the envisioned IWPT scenario, both imaging and WPT functionalities are consolidated within a singular hardware platform, operating on a unified waveform. As these functions utilize the same antenna and transmission power resources, a crucial performance trade-off emerges between imaging and WPT. This trade-off is fundamentally characterized by optimizing the beamforming vector of the transmitted signal within the IWPT system.
The primary contributions of this communication can be succinctly summarized as follows:
\begin{itemize}
\item To our knowledge, we are the first to propose the concept of IWPT, which aims to enhance cost-effectiveness and improve spectrum and energy efficiency by leveraging a shared hardware platform, spectrum, and power resources for two functions. This aligns with the anticipated development trajectory of future 6G networks. In our investigation, we explore two array architectures for this novel concept: the fully digital array and the hybrid array.
\item Additionally, we employ the condition number of the equivalent channel as a metric to evaluate the quality of imaging, providing an effective measure for performance assessment. We analyze the trade-off between imaging and WPT by optimizing the transmit beamforming vectors. Although the beamforming design problem is non-convex, we tackle it iteratively using the semidefinite relaxation approach and successive convex approximation method. Furthermore, for the challenge of coupling digital and analog precoding in the hybrid array, we develop an alternating optimization algorithm to address this issue.
\item We provide numerical results to demonstrate the fundamental performance trade-off between imaging and WPT. Numerical results also verify the effectiveness of the proposed beamforming designs for the novel IWPT system.
\end{itemize}

\textit{Notation:} Scalar variables, vectors and matrices are represented with lower letters, lower bold letters, and capital bold letters, respectively (e.g., $x$, $\bf x$, and $\bf X$, respectively). The term $\mathbb{C}^{N \times N}$ denotes a complex space of dimension ${N \times N}$. $\left( \cdot \right)^H $ denotes the Hermittian operators. For a matrix $\bf R$, ${\bf R} \succeq  {\bf 0}$ means that $\bf R$ is positive semidefinite.

The rest of this paper is organized as follows: Sec.~\ref{sec:Model} introduces the considered array architecture, the signal model and the performance metrics of two functions. In Sec.~\ref{sec:solution}, the design of beamforming design scheme is discussed, while Sec.~\ref{sec:Sims} details numerical simulations. Finally, Sec.~\ref{sec:Conclusions} provides concluding remarks.


\section{System Model}
\label{sec:Model}
In this section, we provide a detailed description of the propsed ISWPT system. We begin by outlining the signal and channel model in Section \ref{subsec:SignalModel}. Following that, in Sec.~\ref{subsec:array}, we introduce the two array architectures utilized throughout this paper. Subsequently, we introduce the imaging and WPT metrics in \ref{subsec:image} and formulate the beamforming design problem for ISWPT systems in \ref{subsec:problem}.


\subsection{Transmitted Signal and Channel Model}
\label{subsec:SignalModel}

\begin{figure}[t!]
\centering
\includegraphics[scale=1]{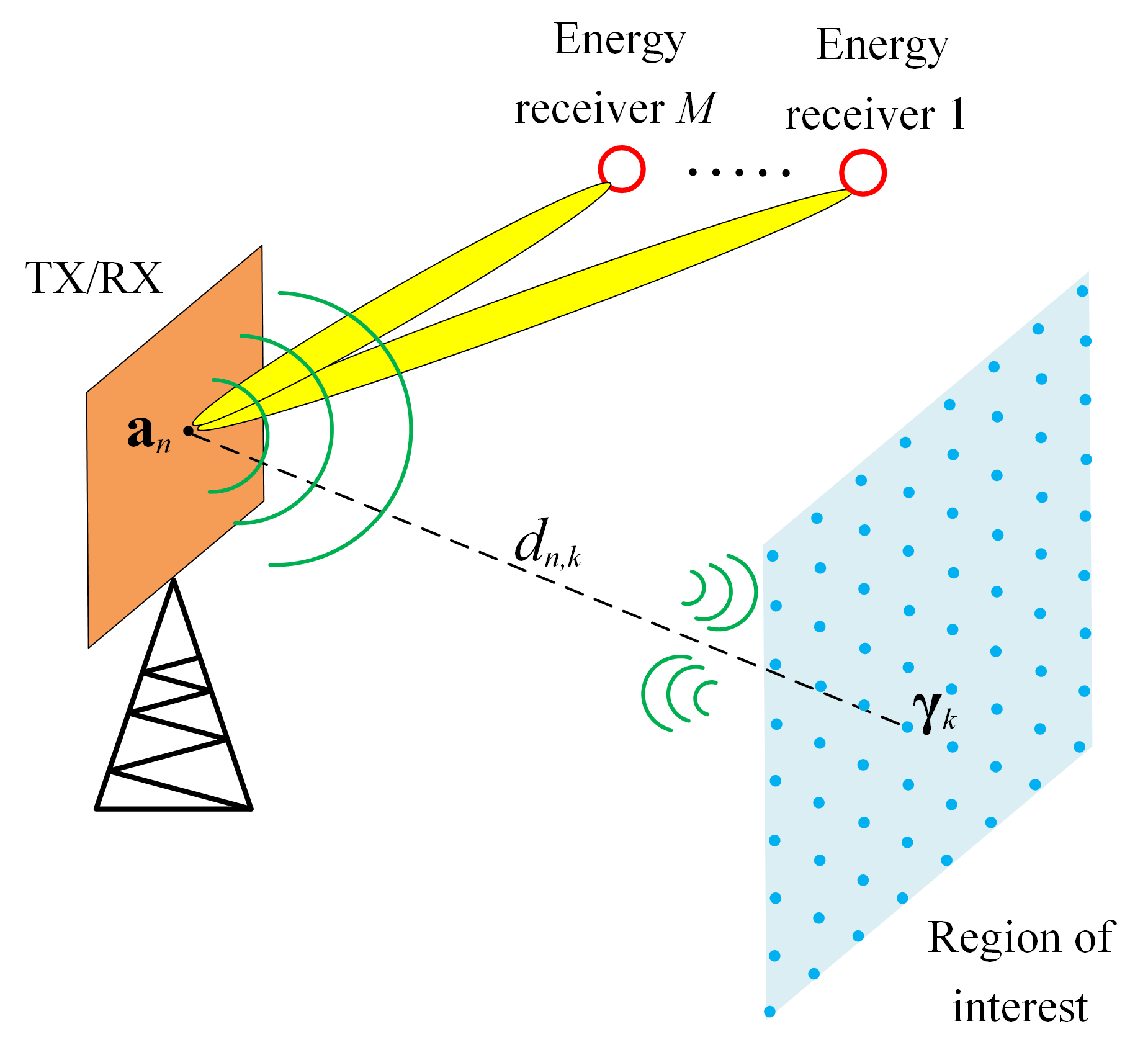}
\caption{Illustration of a joint imaging and WPT scenario.
} 
\label{fig1}
\end{figure}

We consider a joint imaging and wireless power transfer system as illustrated in Fig~\ref{fig1}, wherein the transmitting and receiving discrete antenna arrays (TX/RX) are colocated and equipped with $N$ antennas, the $n$-th antenna is located at ${\bf a}_n = \left[ x^{\rm a}_n, y^{\rm a}_n, z^{\rm a}_n \right], n = 1,2, \cdots, N$. 
The TX illuminates the ROI with the illumination signal and serves $M$ energy receivers simultaneously, where the $m$-th energy receiver is located at ${\bf p}_m = \left[ x^{\rm p}_m, y^{\rm p}_m, z^{\rm p}_m \right], m = 1,2, \cdots, M$. 
The RX observes the reflected signal of the illumination signal from the ROI. The ROI is divided into a grid of $K$ square cells of size $\Delta$, whose positions are ${\bf r}_k = \left[ x^{\rm r}_k, y^{\rm r}_k, z^{\rm r}_k \right], k = 1,2, \cdots, K$. The $k$-th cell is characterized by a scattering coefficient defined as $\gamma_k$, related to the radar cross section (RCS) of the scatterer included in the cell (if any), which have $\text{RCS}_k = |\gamma_k|^2 2\lambda/4\pi$. If the $k$-th cell is empty, then $\gamma_k = 0$. The magnitude of the scattering coefficient $|\gamma_k|$ is upper bounded by the maximum RCS from a scatterer of area $\Delta^2$, which corresponds to the RCS of a perfect electric conductor (PEC) having the same area given by $\text{RCS}_\text{max} = \Delta^2 2\lambda/4\pi$. Therefore, we have $|\gamma_k| \leq \Delta$.

Given that energy receivers can harvest energy from the illumination signal, there is no necessity for designing dedicated WPT signals. In this context, we contemplate employing solely the illumination signal to facilitate both imaging and WPT functionalities. Consequently, the transmitted baseband signal can be represented as ${\bf x} = \left[ x_1, x_2, \cdots, x_N \right]^T \in \mathbb{C}^{N \times 1}$.
To optimize WPT efficiency, we make the assumption that the transmitter (TX) consistently transmits signals at the maximum power level, such that $\Vert {\bf x} \Vert^2 = P_{\rm t}$, where $P_{\rm t}$ signifies the total transmit power budget.

Assuming both imaging and WPT occur in the radiating near-field, i.e., the distance, $d$, between the TX/RX and the energy receiver or ROI should ensure that $d < \frac{2D^2}{\lambda}$, where $D$ and $\lambda$ representing the array diameter and the signal wavelength, respectively. Define ${\bf H}_{\rm T} \in \mathbb{C}^{K \times N}$, ${\bf H}_{\rm R} \in \mathbb{C}^{N \times K}$ and ${\bf G}\in \mathbb{C}^{N \times M}$ as the imaging transmission channel matrix, the imaging receive channel matrix, and the power transmission channel matrix, respectively. Since the TX and RX is colocated, we have ${\bf H}_{\rm R} = {\bf H}^{T}_{\rm T}$.
Then, under the line of sight (LOS) channel condition, the element of the imaging transmission could be expressed as
\begin{equation}
\label{channelGT}
{\bf H}_{\rm T}[k,n] = {\bf H}_{\rm R}[n,k] = F(\Theta_{k,n})\frac{\lambda}{4\pi d_{k,n}}e^{-j\frac{2\pi}{\lambda}d_{k,n}}, 
\end{equation}
and the element of the power transmission channel matrix could be expressed as
\begin{equation}
\label{channelGE}
{\bf G}[m,n] = F(\Theta_{m,n})\frac{\lambda}{4\pi d_{m,n}}e^{-j\frac{2\pi}{\lambda}d_{m,n}}, 
\end{equation}
where $d_{k,n}$ and $d_{m,n}$ are the distance between the corresponding antenna and the ROI cell or the energy receiver, $F(\Theta_{k,n})$ and $F(\Theta_{m,n})$ are the radiation profile, evaluated in the direction of arrival (i.e., $\Theta_{k,n}$ and $\Theta_{m,n}$), which modeled as \cite{pathloss}. For convenience, in the following study, we assume the channel information is known.

\subsection{Array Architecture}
\label{subsec:array}

\begin{figure}
\centering
\subfigure[Fully digital architecture]{
\includegraphics[scale=0.6]{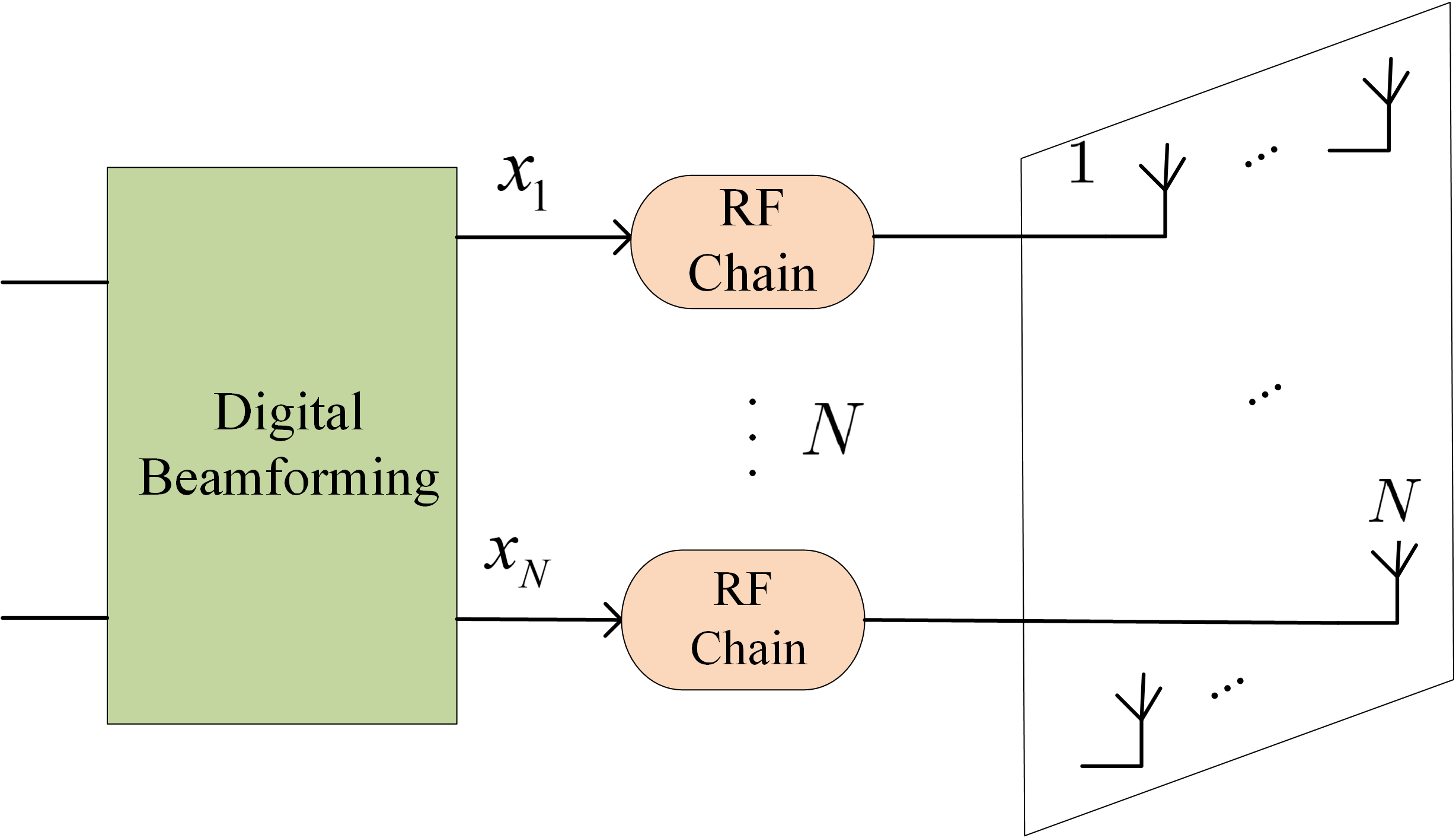}
\label{fig1.1}
}
\subfigure[Partially connected Hybrid architecture]{
\includegraphics[scale=0.6]{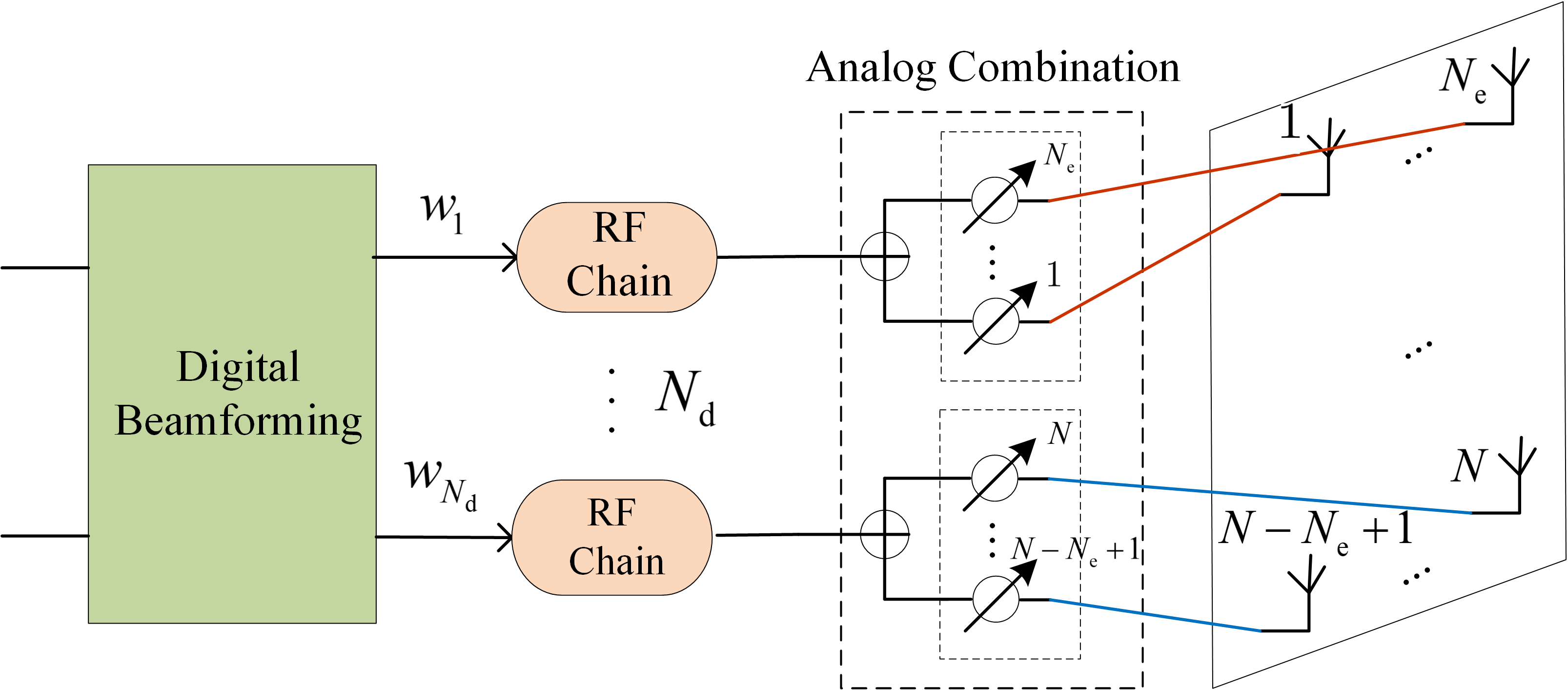}
\label{fig1.2}
}
\caption{Considered array architectures.} 
\vspace{0.4cm}
\label{fig0}
\end{figure}
We consider two types of antenna architectures to generate $\bf x$, depicted in Fig.~\ref{fig0}, that are introduced separately in the following subsections. For convenience, the antennas of all architectures are placed in a rectangular arrangement with $N_{\rm d}$ rows and $N_{\rm e}$ columns, with an overall of $N=N_{\rm d}N_{\rm e}$ elements.

\paragraph{Fully digital architecture} 
The implementation of fully digital antennas, where all processing is conducted in the digital domain, is widely preferred for its remarkable flexibility and the extensive freedom it offers for beamforming \cite{digital}, notwithstanding the potential for increased costs when deploying large-scale arrays.
In architectures employing fully digital antennas, each antenna element is linked to a dedicated Radio Frequency (RF) chain, as shown in Fig.~\ref{fig1.1}. The transmitted baseband signal, denoted as ${\bf x} \in \mathbb{C}^{N \times 1}$, is directly conveyed to the antenna via the RF chain subsequent to digital precoding.

\paragraph{Hybrid architecture}
In our investigation, we center our attention on a partially connected hybrid array configuration, as shown in Figure~\ref{fig1.2}. Within this arrangement, the quantity of Radio Frequency (RF) chains, designated as $N_{\rm d}$, is less than the overall count of antenna elements. Each digital signal output links to an RF chain, and subsequently, each RF chain output connects to the transmit antennas through a partially connected analog beamforming network reliant on phase shifters. This hybrid array architecture adeptly balances complexity and performance, offering a cost-efficient resolution for expansive antenna arrays.
The array is specifically partitioned into $N_{\rm d}$ linear sub-arrays, each consisting of $N_{\rm e}$ elements uniformly arranged. Each linear sub-array is managed by an autonomous phase-shifter network associated with a single RF chain. Consequently, the signal transmitted by the hybrid array is given by
\begin{equation} 
\label{y_h}
{\bf x}={\bf Q} \, {\bf w}. 
\end{equation}
Here ${\bf w}\in \mathbb{C}^{ N_{\rm d} \times 1}$ is the the digital precoding output for the hybrid array, and the matrix ${\bf Q} \in \mathbb{C}^{N_{\rm d} \times N}$ represents analog precoding of the hybrid array that satisfies 
\begin{equation}
\label{stru_h}
{\bf Q}_{n, (i-1) {N_{\rm e}}+l}= \begin{cases}{ q}_{i, l} \in \mathcal{F} & i=n, \\ 0 & i \neq n.\end{cases}
\end{equation}
where ${ q}_{i, l}$ denotes the phase of phase-shifter, which satisfies
\begin{align}
\label{lim_h}
&{ q}_{i, l} \in \mathcal{F} \triangleq\left\{e^{j \phi_{i,l}} \mid \phi_{i,l} \in[0,2 \pi]\right\}, &&\forall i, l.
\end{align}

Note that we did not distinguish the transmission signals of the two array architectures, but unified them as $\bf x$ because they follow the same constraints. The difference is that fully digital antennas can directly design $\bf x$, while hybrid arrays require a joint design scheme for the digital and analog domains to obtain $\bf x$ as \eqref{y_h}. 

\subsection{Imaging and WPT Metric}
\label{subsec:image}

According to the process of the illumination signal reflection as mentioned, the received signal on RX can be expressed as
\begin{equation}
\label{receive}
{\bf y} = {\bf H}_{\rm R} \boldsymbol{\Gamma} {\bf H}_{\rm T}{\bf x} + {\bf n}, 
\end{equation}
where $\boldsymbol{\Gamma} \in \mathbb{C}^{K \times K}$ is a diagonal matrix with the diagonal element $\boldsymbol{\Gamma}_{k,k}=\gamma_k$, ${\bf n}$ is is an additive white Gaussian noise with variance $\sigma^2$.
Let $\boldsymbol{\gamma} = \left[ \gamma_1, \gamma_2, \cdots, \gamma_K\right]^T $, we have $\boldsymbol{\Gamma} = diag(\boldsymbol{\gamma})$, our aim is to estimate $\boldsymbol{\gamma}$ from the received signal, and \eqref{receive} could be rewritten as 
\begin{equation}
\label{receive2}
{\bf y} = {\bf H}_{\rm R} \text{diag} \left( {\bf H}_{\rm T}{\bf x}\right) \boldsymbol{\gamma} + {\bf n}. 
\end{equation}

Without losing generality, we assume that $N>K$, and define ${\bf H} = {\bf H}_{\rm R} \text{diag}\left( {\bf H}_{\rm T}{\bf x}\right)$, ${\bf H}$ can be equivalent regarded as a MIMO channel, then estimate $\boldsymbol{\gamma}$ from \eqref{receive2} becomes a conventional signal recovery problem. Using the common least squares method, the estimated scattering coefficients can be expressed as
\begin{equation}
\label{LS}
\hat{\boldsymbol{\gamma}} = {\bf H}^{\dagger}{\bf y}, 
\end{equation}
where ${\bf H}^{\dagger}$ is the pseudo-inverse of the equivalent channel matrix ${\bf H}$.
Once ${\bf G}_{\rm T}$ and ${\bf G}_{\rm R}$ are fixed, ${\bf H}$ would also determined by the illumination signal, so it is possible to improve the estimation accuracy of $\hat{\boldsymbol{\gamma}}$ in imaging by design $\bf x$.

To enhance the accuracy of estimation in \eqref{LS}, one direct strategy involves scrutinizing the mean square error (MSE), defined as $\mathbb{E}\left\Vert \boldsymbol{\gamma} - \hat{\boldsymbol{\gamma}} \right\Vert^2$. However, estimating $\boldsymbol{\gamma}$ can often present challenges due to ill-posedness \cite{holographic}, particularly when ${\bf H}{\rm T}$ and ${\bf H}{\rm R}$ exhibit rank deficiency in practical near-field scenarios, leading to a rank-deficient composite channel matrix $\bf H$. The potential ill-posed nature of the problem renders the analysis of MSE overly intricate for signal design purposes. Instead, in this paper, we choose the condition number of the equivalent channel as the metric of imaging.

The condition number provides insight into the efficient characterization of the near field MIMO channel, in general, can be defined as
\begin{equation}
\label{cond}
\text{cond} = \frac{\lambda_{max}}{\lambda_{min}}, 
\end{equation}
where $\lambda_{max}$ and $\lambda_{min}$ are the maximum and minimum eigenvalue of {\bf H}, respectively.
When the condition number of a channel approaches 1, the channel is deemed well-conditioned, indicating that its spatial characteristics are in an optimal state. Conversely, a channel tends towards infinite condition numbers when it is rank deficient. Consequently, our objective is to minimize the condition number of the equivalent channel through the design of the illumination signal. This optimization problem can be formulated as
\begin{equation}
\label{op-imag}
\begin{aligned}
&\min _{{\bf x}}~ \text{cond}  \\
&~~s.t.~~ \Vert {\bf x} \Vert^2 = P_{\rm t}.
\end{aligned}
\end{equation}

For WPT operation, the received power of the $m$-th user is contingent on the transmitted signal (here is the illumination signal), which can be represented as 
\begin{equation}
\label{pe}
\begin{aligned}
&P_m \left({\bf R}\right) =\zeta {{\bf g}_m} {\bf R} {\bf g}^H_m,
\end{aligned}
\end{equation}
where $0<\zeta<1$ is the constant energy conversion efficiency, and ${\bf R} = {\bf x}{\bf x}^H$ denotes the covariance matrix of the illumination signal, ${\bf g}_m \in \mathbb{C}^{1 \times N}$ is the $m$-th row of $\bf G$, denotes the wireless channel between the transmitter and the $m$-th energy receiver. 

In general, the goal of WPT is to maximize the sum-harvested power, while adhering to the total transmit power constraint, which can be mathematically formulated as the following maximization problem: 
\begin{equation}
\label{perf_e}
\begin{aligned}
&\max _{{\bf x}}~ \sum_{m=1}^{ M} P_m \left( {\bf R } \right) = \zeta \text{trace} \left({\bf G} {\bf R} {\bf G}^H \right) \\ &~~s.t.~~ {\bf R} = {\bf x}{\bf x}^H, \\ &~~~~~~~~\Vert {\bf x} \Vert^2 = P_{\rm t}.
\end{aligned}
\end{equation}

Following \cite{SWIPT2}, we can obtain a closed-form optimal solution of \eqref{perf_e} directly, which expressed as 
\begin{equation}
\label{WPT}
\begin{aligned}
{\bf x}^\ast = \sqrt{P_{\rm t}}{\bf u}_1 \left( {\bf G}\right),
\end{aligned}
\end{equation}
where ${\bf u}_1 \left( {\bf G}\right)$ denotes the eigenvector corresponding to the maximal eigenvalue of the matrix ${\bf G}$. 


\subsection{Problem Formulation} 
\label{subsec:problem}

The result derived from \eqref{WPT} is improbable to represent the optimal solution of \eqref{op-imag}, indicating the presence of an inherent trade-off between imaging and WPT performance in our systems. In this paper, we would like to characterize this fundamental performance trade-off between imaging and WPT by optimizing the illumination signal ${\bf x}$. Mathematically, our interested beamforming design problem is formulated as 
\begin{equation}
\label{j-op1}
\begin{aligned}
&\min _{{\bf x}}~ \text{cond} \\
&~~s.t.~~ \zeta \text{trace} \left({\bf G} {\bf R} {\bf G}^H \right) \geq E_{\rm r} \\ &~~~~~~~~\Vert {\bf x} \Vert^2 = P_{\rm t}.
\end{aligned}
\end{equation}
Where $E_{\rm r}$ is the set received power threshold that satisfies $E_{\rm r} \in \left[0,E_{max}\right]$, here $E_{max}$ is the achievable maximum received power calculated as $E_{max} = \zeta \text{trace} \left({\bf G} {\bf R}^\ast {\bf G}^H \right)$, and ${\bf R}^\ast$ is the corvirance matrix of ${\bf x}^\ast$ from \eqref{WPT}. 
The extreme case, e.g.,  $E_{\rm r}=0$ or $E_{\rm r}=E_{max}$ corresponds to the scenario of only imaging or only WPT. 

Problem \eqref{j-op1} serves as a comprehensive optimization challenge applicable across all antenna architectures. In the forthcoming section, our focus will revolve around investigating the precoding scheme aimed at acquiring ${\bf x}$ that minimizes \eqref{j-op1} within diverse antenna architectures.


\section{Joint Beamforming Design }
\label{sec:solution}
In this section, we study beam precoding design to achieve the performance trade-off between imaging and WPT. We first explicitly express the relationship between the objective function of our optimization problem and the optimization variables in Sec.\ref{subsec:transefer}. Subsequently, we delve into resolving the aforementioned issue within the framework of a fully digital array in Section \ref{subsec:SDR}, followed by the derivation of the optimal hybrid phase shifter configuration in Section \ref{subsec:hybridop}.


\subsection{Problem Reformulation} 
\label{subsec:transefer}

Problem \eqref{j-op1} is hard to solve due to the absence of a direct relationship between $\bf x$ and the condition number as defined by \eqref{cond}. Inspired by \cite{condnumber}, the problem of minimizing the condition number of ${\bf H}$ can be equivalent to minimizing the trace of the product of ${\bf H}$ and its transpose. This equivalence arises from the fact that minimizing the condition number is akin to minimizing the maximum eigenvalue $\lambda_{max}$. Specifically, in accordance with the definition of the condition number, we have
\begin{equation}
\label{lb-cond}
\text{cond} \leq \left(\frac{\lambda_{max}}{\lambda_{min}} \right)^2. 
\end{equation}
Thus the minimization problem can be regarded as minimize $\lambda_{max}^2$. Further more, for the semidefinite matrix ${\bf H}{\bf H}^H$, we have
\begin{equation}
\label{lb-cond2}
\lambda_{max}^2 < \text{trace} \left( {\bf H}{\bf H}^H \right).
\end{equation}
Based on \eqref{lb-cond} and \eqref{lb-cond2}, problem \eqref{j-op1} can be equivalent formulated as
\begin{subequations}
\label{j-op2}
\begin{align}
&\min _{{\bf x}}~ \text{trace} \left( {\bf H}{\bf H}^H \right) \quad \\
&~~s.t.~~ {\bf H} = {\bf H}_{\rm R} \text{diag}\left( {\bf H}_{\rm T}{\bf x}\right) \label{op2-b} \\ &~~~~~~~~ \zeta \text{trace} \left({\bf G} {\bf R} {\bf G}^H \right) \geq E_{\rm r} \label{op2-c} \\ &~~~~~~~~\Vert {\bf x} \Vert^2 = P_{\rm t}. 
\end{align}
\end{subequations}

In problem \eqref{j-op2}, the optimization variable $\bf x$ is tied to the objective function through \eqref{op2-b}. However, the diagonalization process in \eqref{op2-b} continues to pose challenges in assessing the influence of $\bf x$ on the objective function. To address this issue, we employ a series of matrix operations to eliminate the constraint \eqref{op2-b} in problem \eqref{j-op2}:

Firstly, the objective function of \eqref{j-op2} can be regarded as
\begin{equation}
\label{tran1}
\begin{aligned}
\text{trace} \left( {\bf H}{\bf H}^H \right) = \text{vec} \left( {\bf H} \right)^H \text{vec} \left( {\bf H} \right),
\end{aligned}
\end{equation}
where $\text{vec} \left( \cdot \right)$ denotes the vector stretching operation on matrix, and according to \eqref{op2-b}, we have
\begin{equation}
\label{tran2}
\begin{aligned}
\text{vec} \left( {\bf H} \right) = \left( {\bf I}_N \otimes {\bf H}_{\rm R}\right) \text{vec} \left( \text{diag}\left( {\bf H}_{\rm T}{\bf x}\right) \right).
\end{aligned}
\end{equation}
Substituting \eqref{tran2} into \eqref{tran1}, so we have
\begin{equation}
\label{tran3}
\begin{aligned}
\text{trace} \left( {\bf H}{\bf H}^H \right) &= \text{vec} \left( \text{diag}\left( {\bf H}_{\rm T}{\bf x}\right) \right)^H \left( {\bf I}_N \otimes {\bf H}_{\rm R}\right)^H  ( {\bf I}_N \cdots \\ & ~~~\otimes {\bf H}_{\rm R}) \text{vec} \left( \text{diag}\left( {\bf H}_{\rm T}{\bf x}\right) \right)\\ & \stackrel{(a)}{=} \vert {\bf x}^H {\bf H}_{\rm T}^H {\bf H}_{\rm R}^H {\bf H}_{\rm R}{\bf H}_{\rm T}{\bf x} \vert,
\end{aligned}
\end{equation}
where $\stackrel{(a)}{=}$ comes from that the calculation related to zero elements of $\text{vec} \left( \text{diag}\left( {\bf H}_{\rm T}{\bf x}\right) \right)$ in above equation can be ignored. 

Therefore, the problem \eqref{j-op2} can be rewritten as
\begin{subequations}
\label{j-op3}
\begin{align}
&\min _{{\bf x}}~  \vert {\bf x}^H {\bf H}_{\rm T}^H {\bf H}_{\rm R}^H {\bf H}_{\rm R}{\bf H}_{\rm T}{\bf x} \vert   \\
&~~s.t.~~ \zeta \text{trace} \left({\bf G} {\bf R} {\bf G}^H \right) \geq E_{\rm r}  \label{op3-b}\\ &~~~~~~~~ \Vert {\bf x} \Vert^2 = P_{\rm t}. 
\end{align}
\end{subequations}

\subsection{Optimal Digital Beamforming Design}
\label{subsec:SDR}

Problem \eqref{j-op3} is non-convex due to the quadratic equality constraint \eqref{op3-b}, rendering it intricate to solve directly. To tackle this issue, we use the SDR approach \cite{luo2010semidefinite} to transform \eqref{j-op3}. According to the definition of $\bf R$, \eqref{j-op3} can be rewritten as an equivalent quadratic semidefinite programming (QSDP) problem with rank-1 constraints:
\begin{equation}
\label{j-op4}
\begin{aligned}
&\min _{{\bf R}}~  \text{trace} \left( {\bf H}_{\rm T}^H {\bf H}_{\rm R}^H {\bf H}_{\rm R}{\bf H}_{\rm T}{\bf R}  \right)  \\
&~~s.t.~~ \zeta \text{trace} \left({\bf G} {\bf R} {\bf G}^H \right) \geq E_{\rm r}  \\ &~~~~~~~~ \text{trace}({\bf R}) = P_{\rm t},  \\ &~~~~~~~~ \text{rank}({\bf R}) = 1, \\ &~~~~~~~~{\bf R} \succeq {\bf 0}.
\end{aligned}
\end{equation}

Due to the non-convex rank-one constraint, the problem \eqref{j-op4} is still a non-convex optimization problem. Next, we use the difference-of-convex (DC) programming to handle this rank-one constraint. Specifically, for the semi-definite matrix $\bf R$, the rank-1 constraint can be equivalent expressed as 
\begin{equation}
\label{rank1}
\begin{aligned}
\text{trace}({\bf R}) = \left\Vert {\bf R} \right\Vert_2.
\end{aligned}
\end{equation}
Therefore, we can transform the rank-1 constraint of ${\bf R}$ into a penalty term to the objective function of the problem \eqref{j-op4}, and the transformed problem is formulated as
\begin{equation}
\label{j-op5}
\begin{aligned}
&\min _{{\bf R}}~  \text{trace} \left( {\bf H}_{\rm T}^H {\bf H}_{\rm R}^H {\bf H}_{\rm R}{\bf H}_{\rm T}{\bf R} \right) \\&~~~~~~+ \eta \left( \text{trace}({\bf R}) - \left\Vert {\bf R} \right\Vert_2 \right)  \\
&~~s.t.~~ \zeta \text{trace} \left({\bf G} {\bf R} {\bf G}^H \right) \geq E_{\rm r}  \\ &~~~~~~~~ \text{trace}({\bf R}) = P_{\rm t},   \\ &~~~~~~~~ {\bf R} \succeq {\bf 0}.
\end{aligned}
\end{equation}
where $\zeta \gg 0$ is a penalty factor related to the rank-1 constraint. \eqref{j-op5} is still non-convex due to the presence of a 2-norm $\left\Vert {\bf R} \right\Vert_2$ in the objective function, so we utilize the SCA method to write a first order Taylor expansion of $\left\Vert {\bf R} \right\Vert_2$ as its lower bound:
\begin{equation}
\label{approx}
\begin{aligned}
\left\Vert {\bf R} \right\Vert_2 &\geq \left\Vert {\bf R}^{(t)} \right\Vert_2 + \text{trace}\left( {\bf u}_{max}^{(t)} {\bf u}_{max}^{(t)H} ({\bf R} - {\bf R}^{(t)}) \right) \\& \triangleq f\left( {\bf R}^{(t)}, {\bf R} \right),
\end{aligned}
\end{equation}
where ${\bf R}^{(t)}$ is the feasible point of the first order Taylor expansion in the $t$-th iteration, ${\bf u}_{max}^{(t)}$ is the eigenvector corresponding to the largest singular value of the matrix ${\bf R}^{(t)}$. Hence, by replacing $\left\Vert {\bf R} \right\Vert_2$ with $f\left( {\bf R}^{(t)}, {\bf R} \right)$ in the $t$-th iteration, \eqref{j-op5} can be approximately transformed into
\begin{equation}
\label{j-op6}
\begin{aligned}
&\min _{{\bf R}}~  \text{trace} \left( {\bf H}_{\rm T}^H {\bf H}_{\rm R}^H {\bf H}_{\rm R}{\bf H}_{\rm T}{\bf R} \right) \\ &~~~~~~+ \eta \left( \text{trace}({\bf R}) - f( {\bf R}^{(t)}, {\bf R} ) \right)  \\
&~~s.t.~~ \zeta \text{trace} \left({\bf G} {\bf R} {\bf G}^H \right) \geq E_{\rm r}  \\ &~~~~~~~~ \text{trace}({\bf R}) = P_{\rm t},   \\ &~~~~~~~~ {\bf R} \succeq {\bf 0}.
\end{aligned}
\end{equation}

${\bf R}^{(t)}$ is iteratively updated. In each iteration, ${\bf R}^{(t)}$ is taken from the solution of problem \eqref{j-op6} in the previous iteration, and the resulting ${\bf R}^{(t)}$ would iteratively close to the optimal solution of \eqref{j-op5}. Meanwhile, we set that ${\bf R}^{(1)} = {\bf R}^\ast$ to ensure the initial satisfies all constraints of \eqref{j-op6}.
The resulting problem \eqref{j-op6} is a typical QSDP problem, which is convex and can be solved directly using existing convex optimization solvers such as CVX \cite{grant2014cvx}. 

Once ${\bf R}^{(t)}$ reaches convergence, it can be approximated as a rank 1 matrix, i.e., the optimal solution of problem \eqref{j-op4} would be found. The jointly optimal beam ${\bf x}^{opt}$ for imaging and WPT could be calculated from the eigendecomposition of ${\bf R}^{(t)}$ directly, which expressed as   
\begin{equation}
\label{opt-x}
\begin{aligned}
{\bf x}^{\star} = \lambda_{max}^{(t)} {\bf u}_{max}^{(t)},
\end{aligned}
\end{equation}
where $\lambda_{max}^{(t)}$ is the largest singular value of the matrix ${\bf R}^{(t)}$, since the transmit power constraint, we have $\lambda_{max}^{(t)} = \sqrt{P_{\rm t}}$, thus \eqref{opt-x} could be also expressed as ${\bf x}^{\star} = \sqrt{P_{\rm t}} {\bf u}_{max}^{(t)}$. 

The worst case complexity to solve the QSDP problem \eqref{j-op6} is $\mathcal{O} \left( N^{6.5}  \log (1 / \epsilon) \right)$ with the primal-dual interior-point algorithm \cite{complex}, where $\epsilon$ is the solution accuracy. We assume that the number of iterations required for the SCA process to reach convergence is $K$, thus the whole complexity of the beamforming design is expressed as $\mathcal{O} \left( K N^{6.5}  \log (1 / \epsilon) \right)$.

\subsection{Joint Analog and Digital Design for Hybrid array}
\label{subsec:hybridop}

As outlined in Section \ref{subsec:array}, in the scenario of a fully digital array, the optimal beamforming design can be directly acquired by addressing problem \eqref{j-op1}. Conversely, in the context of a hybrid array, considering \eqref{y_h}, the beamforming design predicament necessitates a collective precoding design spanning the digital and analog domains. Analogous to \eqref{j-op4}, the optimization conundrum \eqref{j-op1} under the hybrid array scenario can be reformulated as
\begin{equation}
\label{j-hop1}
\begin{aligned}
&\min _{{\bf Q},{\bf w}}~  \vert {\bf w}^H{\bf Q}^H {\bf T} {\bf Qw} \vert \\
&~~s.t.~~ \zeta \text{trace} \left({\bf G} {\bf R} {\bf G}^H \right) \geq E_{\rm r} \\ &~~~~~~~~\Vert {\bf Qw} \Vert^2 = P_{\rm t} \\ &~~~~~~~~\eqref{stru_h}, \eqref{lim_h}.
\end{aligned}
\end{equation}
Where ${\bf T} = {\bf H}_{\rm T}^H {\bf H}_{\rm R}^H {\bf H}_{\rm R}{\bf H}_{\rm T}$, and ${\bf R} = {\bf Qw}{\bf w}^H{\bf Q}^H$. 

Since Problem \eqref{j-hop1} involves coupled optimization variables and unit modulus constraints, it is non-convex.
To address this, we aim to derive a hybrid solution for Problem \eqref{j-hop1} that closely approximates the fully-digital solution of another problem, denoted as Problem \eqref{j-op4}. This approach is quite often employed for optimizing hybrid analog/digital systems. Essentially, our objective is to determine a hybrid precoding mapping that closely resembles the fully-digital precoding in the context of the Frobenius norm.

Specifically, let ${\bf x}^{\star} \in \mathbb{C}^{N \times 1}$ be the result of the optimal digital beamforming design by solving problem \eqref{j-op6}. The resulting surrogate optimization problem is given by
\begin{equation}
\label{j-hop2}
\begin{aligned}
&\min _{{\bf Q},{\bf w}}~  \Vert {\bf x}^{\star} - {\bf Qw} \Vert^2 \\
&~~s.t.~~ \Vert {\bf w} \Vert^2 = \frac{P_{\rm t}}{N_{\rm d}} \\ &~~~~~~~~\eqref{stru_h}, \eqref{lim_h}.
\end{aligned}
\end{equation}
Where the power constraint in \eqref{j-hop2} comes from the $\Vert {\bf Qw} \Vert^2 = P_{\rm t}$ in \eqref{j-hop1}, according to \eqref{stru_h} and \eqref{lim_h}, we have $\Vert {\bf Qw} \Vert^2 = \text{trace}({\bf w}^H{\bf Q}^H{\bf Qw}) = N_{\rm d}\text{trace}({\bf w}^H{\bf w})$.

We tackle \eqref{j-hop2} using the alternating optimization. In particular, for a given ${\bf Q}$, the digital precoding result ${\bf w}^{\star}$ which minimizes \eqref{j-hop2} is stated in the following least-squares solution:
\begin{equation}
\label{opt-w}
\begin{aligned}
{\bf w}^{\star} = \left( {\bf Q}^H {\bf Q}\right)^{-1} {\bf Q}^H {\bf x}^{\star} = {\bf Q}^H {\bf x}^{\star}/N_{\rm d},
\end{aligned}
\end{equation}
the least-squares solution in \eqref{opt-w} dose not satisfy the power constraint in \eqref{j-hop2}. Nonetheless, once ${\bf w}$ and ${\bf Q}$ are tuned to optimize \eqref{j-hop2}, we can update the digital precoder by multiplying a factor, i.e., ${\bf w}^{\star} = \frac{\sqrt{P_{\rm t}}{\bf w}^{\star}}{\sqrt{N_{\rm d}}\Vert{\bf w}^{\star} \Vert}$.

When optimizing ${\bf Q}$ with a fixed ${\bf w}$, we leverage the partially connected structure constraint \eqref{stru_h} and the unit-modulus constraint \eqref{lim_h} on the matrix ${\bf Q}$. Through these constraints, a closed-form solution for ${\bf Q}$ emerges. Specifically, when ${\bf w}$ is given, the phase shifter coefficient matrix ${\bf Q}$ that minimizes \eqref{j-hop2} can be attained as per the following proposed theorem.

\begin{theorem}
 \label{thm:opQ}
For a given ${\bf w}$, the matrix ${\bf Q}$ which minimizes \eqref{j-hop2} is
\begin{equation}
\label{opt-Q}
\begin{aligned}
{\bf Q}^{\star}_{n, (i-1) {N_{\rm e}}+l}= \begin{cases}{ q}^{\star}_{i, l} \in \mathcal{F} & i=n, \\ 0 & i \neq n.\end{cases}
\end{aligned}
\end{equation}
here ${q}^{\star}_{i, l}$ is the corresponding phase shifter coefficient which given by 
\begin{equation}
\label{opt-q}
\begin{aligned}
{ q}^{\star}_{i, l} = e^{j\left( \angle {x}^{\star}_{(i-1) {N_{\rm e}}+l} - \angle {\bf w}_{i} \right)}.
\end{aligned}
\end{equation}
where $\angle \cdot$ denotes the angle of complex value.

\end{theorem}
\begin{IEEEproof} 
The proof is given in Appendix~\ref{app:Proof0}.
\end{IEEEproof}

The result of Theorem.~\ref{thm:opQ} indicates that the joint precoding result of the partially connected hybrid array can not perfectly match the result of a fully digital scheme. This discrepancy arises due to the insufficient number of RF chains to impeccably regulate the signal amplitudes at the antennas within the context of the considered partially connected architecture. This disparity isn't a consequence of the chosen scheme but rather stems from the inherent distortion induced by the structural constraints of partially connected hybrid arrays. Consequently, this mismatch engenders a suboptimal trade-off in performance for the hybrid array concerning IWPT compared to the fully digital array.

Finally, the overall resulting configuration algorithm for joint digital and analog precoding for the hybrid array is summarized as the following algorithm.

\begin{algorithm}
\caption{Alternating Optimization of Digital and Analog Precoding for Hybrid Array}
\label{alg:Alternating}
\begin{algorithmic}[1] 
\renewcommand{\algorithmicrequire}{\textbf{Initialize:}} 
\Require Initialize the matrix ${\bf Q}^0 \in \mathbb{C}^{N_{\rm d} \times N}$ constrained by the array architecture with random phase set; Iterations limit $T_{max}$.
\State Obtain ${\bf x}^{\star}$ by solving problem \eqref{j-op6}.
\renewcommand{\algorithmicrequire}{\textbf{While} {$t \leq  T_{max}$} \textbf{do}} \Require  
    \State Calculate ${\bf w}^{t}$ as \eqref{opt-w} with the given matrix ${\bf Q}^{t-1} $ and update as ${\bf w}^{t} = \frac{\sqrt{P_{\rm t}}{\bf w}^{t}}{\sqrt{N_{\rm d}}\Vert{\bf w}^{t} \Vert}$.
    \State Update the coefficient ${ q}^{t}_{i,l} $ as \eqref{opt-q} with the given ${\bf w}^{t}$.  
    \State Update the matrix ${\bf Q}^{t}$ as \eqref{opt-Q}. 
    \State $t=t+1$
\renewcommand{\algorithmicrequire}{\textbf{End while}} \Require
\renewcommand{\algorithmicrequire}{\textbf{Output:}} \Require ${\bf Q}={\bf Q}^{T_{max}}$; ${\bf w}={\bf w}^{T_{max}}$.
\end{algorithmic}
\end{algorithm}

\begin{figure*} 
  \centering 
  \subfigure[Original image]{ 
    \label{fig:original}
    \includegraphics[width = 0.25\linewidth]{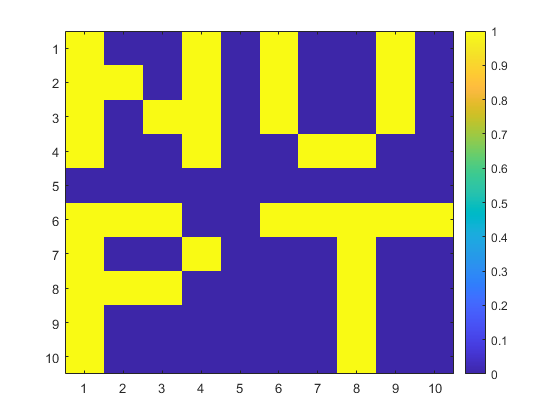} 
  }    
  \subfigure[Random]{ 
    \label{fig:random} 
    \includegraphics[width = 0.25\linewidth]{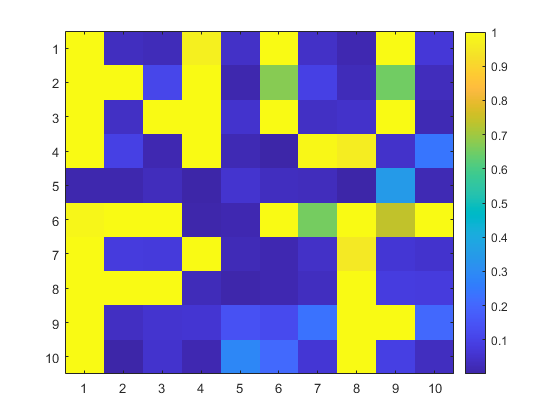} 
  } 
  \subfigure[Imaging only]{ 
    \label{fig:imagonly}
    \includegraphics[width = 0.25\linewidth]{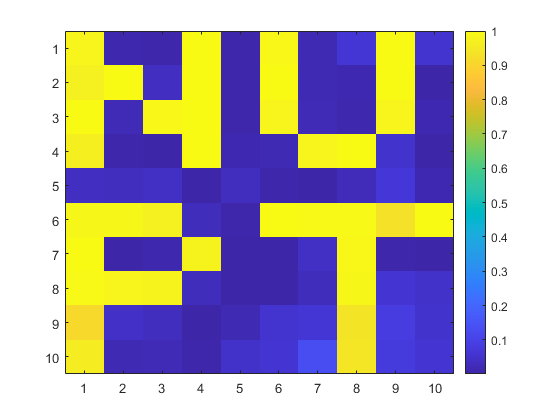} 
  }
  \subfigure[WPT only]{ 
    \label{fig:WPT} 
    \includegraphics[width = 0.25\linewidth]{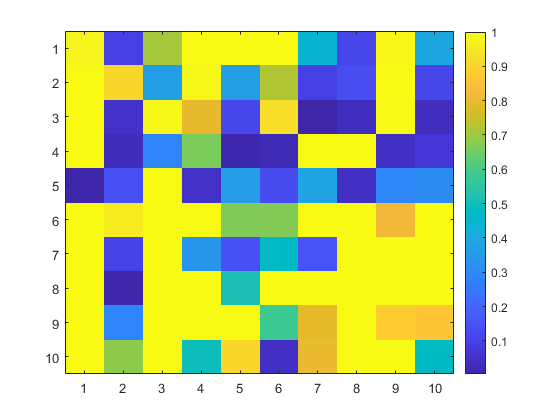} 
  }
  \subfigure[IWPT with fully digital array]{ 
    \label{fig:jointly} 
    \includegraphics[width = 0.25\linewidth]{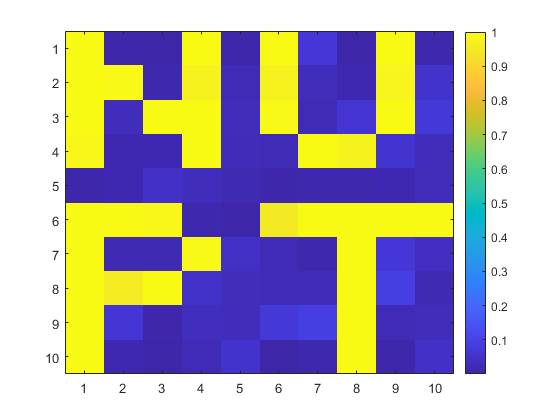} 
  }
  \subfigure[IWPT with hybrid array]{ 
  \label{fig:hybrid}
  \includegraphics[width = 0.25\linewidth]{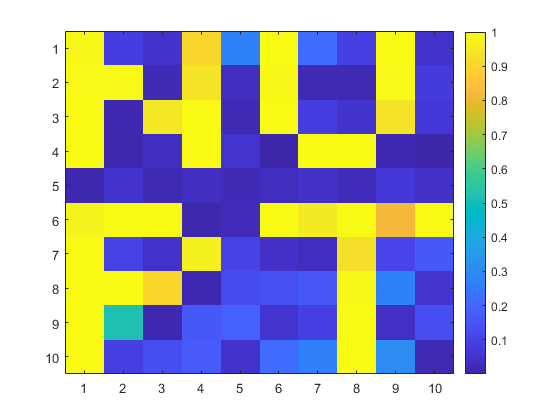} 
  }
  \caption{Imaging results in different illumination design schemes: (a) Original pattern; (b) Random; (c) Imaging only; (d) WPT only; (e) IWPT with fully digital array, where the power threshed is $E_{\rm r} = 0.15E_{max}$; (f) IWPT with hybrid array, where $E_{\rm r} = 0.15E_{max}$.} 
  \label{fig2} 
\end{figure*}

In Algorithm~\ref{alg:Alternating}, it can be seen that ${\bf Q}$ only needs one update to achieve phase matching based on Theorem.~\ref{thm:opQ}, while the update of ${\bf w}$ is only related to ${\bf Q}$ by \eqref{opt-w}, so the algorithm only needs two iterations to converge.
In addition, since the update process only involves closed form solution calculation, the complexity of Algorithm~\ref{alg:Alternating} mainly comes from the calculation of ${\bf x}^{\star}$, which be expressed as $\mathcal{O} \left( K N^{6.5}  \log (1 / \epsilon) \right)$.

%

\section{Numerical Evaluations}
\label{sec:Sims}

\subsection{Simulation Parameters} \label{parameter}
In this section, we provide numerical results to demonstrate the fundamental trade-off between imaging and WPT under a novel IWPT scenario. 
In our experiments, the TX/RX is set at configured as uniform squared arrays, with the array's reference position situated at $(0,0,0)$ m on the Y Z-plane. The region of ROI forms a rectangle centered at $(2,0,0)$ m and is parallel to the Y Z-plane. We direct the TX energy towards three power users concurrently, positioned at $(1.5, 1, 1)$ m, $(1, -1.5, 0)$ m, $(1.5, -1, 0)$ m, respectively.Given these specifications, the maximum received power for WPT can be computed as $E_{\text{max}}$. Furthermore, the other simulation parameters are given in Table~\ref{tab:experiment}.
\begin{table}[t]
\centering
\caption{Experiment Parameters}
\begin{tabular}{|c|c|}
\hline
\textbf{Parameter}                 & \textbf{Value}                             \\ \hline
Transmission bandwidth             & $\Delta f = 120$ kHz                       \\ \hline
Central frequency                  & $f_c = 28$ GHz                             \\ \hline
Wavelength                         & $\lambda = 0.01$ m                         \\ \hline
Noise power spectral density       & $\sigma^2 = -170$ dBm/Hz                   \\ \hline
Transmitted power                  & $P_{\rm t} = 30$ dBm                       \\ \hline
Energy conversion efficiency       & $\zeta = 0.5$                              \\ \hline
Array antenna row                & $N_{\rm d} = 13$ \\ 
                    \hline
Array antenna column                & $N_{\rm e} = 13$ \\ 
                     \hline
Antenna Spacing                      & $3\lambda/2$                               \\ \hline
Antenna Configuration              & $13 \times 13$ antennas                    \\ \hline
Cell Distribution of ROI                 & $10 \times 10$ cells   \\ \hline
Cell Size                  & $\Delta = 0.1$ m    \\ \hline
TX/RX Near Field Distance             & $d_F = 6.5$ m                              \\ \hline
\end{tabular}
\label{tab:experiment}
\end{table}

The numerical indicators of imaging performance are represented by root mean square error (RMSE), calculated as
\begin{equation}
\label{RMSE}
\begin{aligned}
\text{RMSE} = \sqrt{\frac{1}{N_{\rm t}} \sum_{t = 1}^{N_{\rm t}}\Vert \hat{\boldsymbol{\gamma}}^{(t)} - \boldsymbol{\gamma} \Vert^2 },
\end{aligned}
\end{equation}
where $N_{\rm t}$ is the number of Monte Carlo experiments, which set as $N_{\rm t} = 200$ in this paper, and $\hat{\boldsymbol{\gamma}}^{(t)}$ is the estimate in the $t$-th experiment.

\subsection{Numerical Results}
\label{ssec:results}
To assess the imaging efficiency of the proposed beamforming scheme, we initially establish the scenario where $\gamma_k = \Delta$ when the $k$-th cell of the Region of Interest (ROI) contains a scatter to be detected; otherwise, $\gamma_k = 0$. The image under consideration for estimation is illustrated in Fig.~\ref{fig:original}. Our objective is to estimate the scattering coefficients $\boldsymbol{\gamma}$, which represent the original image, under various illumination signal beamforming strategies.

In our results, following cases are considered: 1) \textit{Random:} In this scenario, the illumination signal is generated as a random signal, serving as a fundamental performance benchmark. The estimated outcome is depicted in Fig.~\ref{fig:random}; 2) \textit{Imaging only:} The beamforming process focuses solely on optimizing imaging performance. The corresponding illumination is derived by solving \eqref{j-op6} with $E_{\rm r}=0$. The resulting estimate is illustrated in Fig.~\ref{fig:imagonly}; 3) \textit{WPT only:} Here, the illumination signal is directly obtained via \eqref{WPT}. The beamforming aims at maximizing the total received power of the power users exclusively. The estimated outcome is shown in Fig.~\ref{fig:WPT}; 4) \textit{IWPT with fully digital array:} The optimal signal beam for imaging and Wireless Power Transfer (WPT) is determined by solving \eqref{j-op6} from the proposed beamforming design scheme, with the power threshold set at $E_{\rm r} = 0.15E_{\text{max}}$. The estimated result is displayed in Fig.~\ref{fig:jointly}; 5) \textit{IWPT with hybrid array:} The signal beam is computed using \eqref{y_h}, where the digital and analog precoding are obtained from the proposed Algorithm~\ref{alg:Alternating}.

The random illumination scheme demonstrates a fundamental ability to recover the original image from the received signal overall, albeit with some residual distortion. In contrast, the imaging-only scenario exhibits the most effective image restoration, showcasing how optimizing the equivalent channel conditions can significantly enhance imaging performance, while the WPT-only estimation yields the most distorted results, underscoring the impact of WPT considerations on imaging fidelity. Furthermore, the estimation outcomes of the IWPT scenario with a fully digital array closely resemble those of the imaging-only scenario. This alignment underscores the efficacy of our proposed beamforming optimization design. Conversely, the performance of the IWPT scheme with the hybrid array falls between that of the fully digital setup and the random scheme. This placement corroborates our earlier analysis in Sec.~\ref{subsec:hybridop}, but it should be emphasized that despite this, the hybrid array outperformed the random scheme while significantly cutting down the required number of RF chains by a factor of 13 compared to the all-digital array.

\begin{figure} 
  \centering 
  \subfigure[Accessible condition number versus received power.]{ 
    \label{fig:c-p}
    \includegraphics[width = 0.95\linewidth]{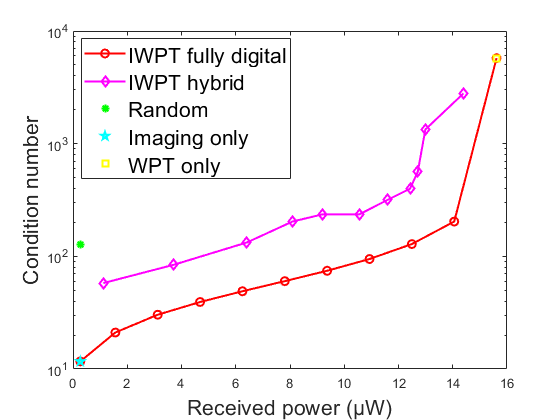} 
  } 
  \subfigure[RMSE versus received power.]{ 
    \label{fig:r-p} 
    \includegraphics[width = 0.95\linewidth]{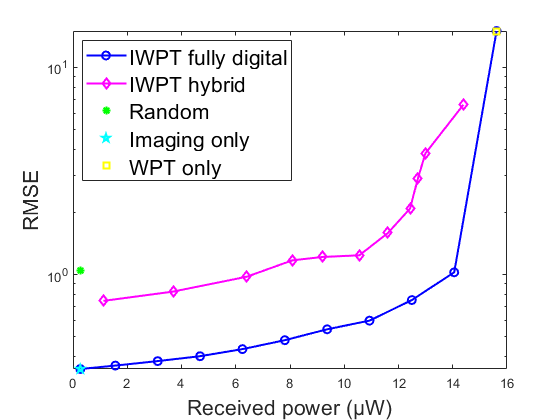} 
  }
    
  \caption{Trade-off between iamging and WPT under different power threshold: (a) Condition number - received power; (b) RMSE - received power.} 
  \label{fig3} 
\end{figure}

In Fig.\ref{fig3}, we illustrate the trade-off between imaging and Wireless Power Transfer (WPT). For detailed insights, Fig.\ref{fig:c-p} and Fig.~\ref{fig:r-p} depict the correlation between the condition number and received power, as well as the Root Mean Square Error (RMSE) and received power, respectively. The comparison points "Random," "Imaging only," and "WPT only" are derived from averaging 200 Monte Carlo experiments within the mentioned context. It is evident that both the condition number of the equivalent channel and the RMSE of imaging increase as the received power of users rises, indicating a fundamental trade-off between imaging and WPT. Enhancing performance in one aspect typically comes at the expense of the other. Our proposed joint optimal scheme can achieve two extremes—the imaging-only scenario and the WPT-only scenario—and offers a flexible trade-off by adjusting the power threshold. As previously discussed, the performance of the hybrid array setup is suboptimal compared to the fully digital array case, with noticeable fluctuations in the lines as the power threshold increases. This variability occurs as the trade-off gradually shifts towards WPT, suggesting that a certain level of distortion may benefit imaging and vice versa. Nonetheless, this phenomenon does not alter the overarching trend of the trade-off performance.

The comparison between Fig.~\ref{fig:c-p} and Fig.~\ref{fig:r-p} reveals a positive correlation between the condition number and the Root Mean Square Error (RMSE) of imaging, particularly noticeable in the hybrid array scenario where the RMSE results exhibit similar fluctuations as the condition number. Despite a non-linear relationship, the RMSE results are indirectly influenced by optimizing the condition number. Given the complexity introduced by the near-field environment, analyzing imaging RMSE directly can be intricate, making the condition number a simpler yet effective optimization target. 
Furthermore, our joint optimization strategy based on imaging and WPT indicates that even under high power constraints, the imaging performance remains superior to that of random illumination. In Fig.~\ref{fig:c-p}, the power gating for improved imaging surpasses 12.2 $\mu$W, while in Fig.~\ref{fig:r-p}, it reaches 14 $\mu$W, showcasing the substantial potential of combining imaging with WPT.

\begin{figure}[t!]
\centering
\includegraphics[width = 0.95\linewidth]{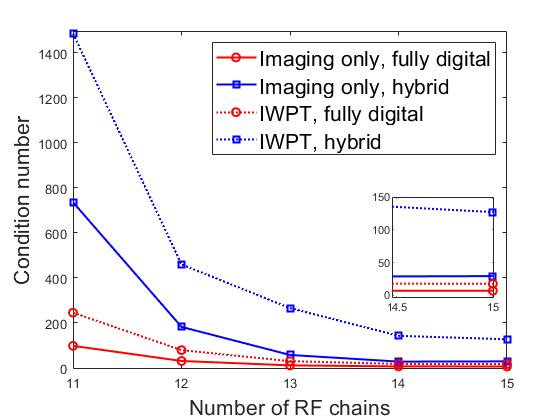}
\caption{Condition number contrast under different RF chain numbers, in IWPT case, the power threshed is set to $E_{\rm r} = 0.15E_{max}$.
} 
\label{fig5}
\end{figure}

The hybrid array design achieves a reduction in RF chains at the expense of a certain performance trade-off. In Fig.\ref{fig5}, we delve into the correlation between the number of RF chains in the hybrid array and the condition number. Keeping other simulation parameters consistent with Sec.~\ref{parameter}, we vary the number of antenna rows for TX/RX , i.e., $N_{\rm d}$, and calculate the corresponding condition number. As $N_{\rm d}$ changes, the total antenna count $N$ also changes, allowing us to compare the results with a fully digital array setup having an identical antenna configuration, the later requiring $13\times$ more RF chains than the hybrid array. We examine two scenarios: imaging only (i.e., the power threshold $E_{\rm r}=0$) and IWPT with $E_{\rm r} = 0.15E_{\rm max}$. The trend reveals that the condition number decreases with an increase in RF chains (or antenna numbers in the fully digital array), leading to a gradual reduction in the performance gap between the hybrid and fully digital arrays to a stable level. This suggests that the performance of the hybrid array hinges on that of the fully digital array it emulates, with a persistent performance gap due to RF chain reduction. 
The performance gap between hybrid arrays and fully digital arrays is narrower in the imaging-only scenario compared to the IWPT scenario. This difference is a consequence of the inherent trade-off between imaging and WPT, which exacerbates the mismatch when aligning the imaging beam with the hybrid array.

\section{Conclusions}
\label{sec:Conclusions}

In this paper, we introduced the concept of Integrated Imaging and Wireless Power Transfer (IWPT), which combines imaging and wireless power transfer functionalities on a unified hardware platform. IWPT utilizes a transmitting array to efficiently illuminate a specific Region of Interest (ROI), enabling the extraction of the ROI's scattering coefficients while simultaneously providing wireless power to nearby users. We characterized the fundamental trade-off between imaging and wireless power transfer of IWPT system by optimizing the illumination signals, under both the fully digital array and hybrid array architectures. With imaging operating in the near-field, we formulated the illumination signal design as an optimization problem aimed at minimizing the condition number of the equivalent channel. To solve this optimization problem, we proposed a semi-definite relaxation-based approach for the fully digital array and an alternating optimization algorithm for the hybrid array. Our numerical results validate the effectiveness of our proposed solutions and highlight the trade-off between imaging and wireless power transfer. These findings underscore the potential of IWPT systems for practical applications, demonstrating their capability to enhance imaging precision and power transfer efficiency in future wireless networks.


\begin{appendices}
	\numberwithin{proposition}{section} 
	\numberwithin{lemma}{section} 
	\numberwithin{corollary}{section} 
	\numberwithin{remark}{section} 
	\numberwithin{equation}{section}

		\vspace{-0.2cm}
	\section{Proof of Theorem \ref{thm:opQ}}
	\label{app:Proof0}	

To drop the non-convex structure constraint \eqref{stru_h}, we define that ${\bf q} = \text{vec} \left( {\bf Q} \right)$ and ${\bf V} = {\bf w}^T \otimes {\bf I}_N$. The objective function of \eqref{j-hop2} can be rewritten as
\begin{equation}
\label{proof1}
\begin{aligned}
\Vert {\bf x}^{\star} - {\bf Qw} \Vert^2 = \Vert {\bf x}^{\star} - \overline{{\bf V} }\overline{{\bf q}} \Vert^2
\end{aligned}
\end{equation}
where $\overline{\bf q}$ is the modified version of ${\bf q}$ obtained by removing all the zero elements of ${\bf q}$ and thus $\overline{\bf q}\in \mathbb{C}^{N \times 1}$, $\overline{\bf V}$ is the modified version of ${\bf V}$ obtained by removing the elements having the same index as the zero elements of ${\bf q}$. By substituting \eqref{proof1} into \eqref{j-hop2} with given ${\bf w}$, \eqref{j-hop2} can be rewritten as a problem about optimizing $\overline{\bf q}$ which given by
\begin{equation}
\label{proof2}
\begin{aligned}
&\min _{\overline{\bf q}}~  \Vert {\bf x}^{\star} - \overline{{\bf V} }\overline{{\bf q}} \Vert^2 \\
&~~s.t.~~ \overline{q}_{n} \in \mathcal{F}.
\end{aligned}
\end{equation}
where $\overline{q}_{n}$ denotes the $n$-th element of $\overline{\bf q}$. In addition, according to the definition ${\bf V}$, we obverse that $\overline{\bf V}$ is a diagonal matrix with the diagonal elements vector expressed as $[w_1 {\bf l}_{N_{\rm e}^T}, \cdots, w_{N_{\rm d}} {\bf l}_{N_{\rm e}}^T]^T$.
Therefore, \eqref{proof2} can be futher decomposed into $N_{\rm d}$ subproblems, each with an identical structure. In particular, each subproblem individually designs the weighting coefficients for a single RF chains, with the $i$-th subproblem in scalar form given by
\begin{equation}
\label{proof3}
\begin{aligned}
&\min _{\overline{ q}_{n}}~  \vert { x}_n^{\star} - w_i \overline{{ q}}_n \vert^2 \\
&~~s.t.~~ \overline{q}_{n} \in \mathcal{F}, \\
&~~~~~~~~n = (i-1)N_{\rm e},\cdots,iN_{\rm e}.
\end{aligned}
\end{equation}

Since $\overline{ q}_{n}$ is only related to the phase, the solution to problem \eqref{proof3} is $\overline{ q}_{n} = e^{j\left( \angle {x}^{\star}_{n} - \angle {\bf w}_{i} \right)}$.
Therefore, according to the mapping relationship between $\overline{q}_{n}$ and $q_{i,l}$, i.e., $q_{i,l} = \overline{q}_{(i-1)N_{\rm e}+l}$, the optimal phase shifter coefficient ${q}^{\star}_{i, l}$ is given by
\begin{equation}
\label{proof4}
\begin{aligned}
{ q}^{\star}_{i, l} = e^{j\left( \angle {x}^{\star}_{(i-1) {N_{\rm e}}+l} - \angle {\bf w}_{i} \right)}.
\end{aligned}
\end{equation}
This proves the theorem.

\end{appendices}

	\bibliographystyle{IEEEtran}
	\bibliography{IEEEabrv,refs}

\end{document}